\documentstyle[12pt,cmp209,epsf,wrapfig]{article}
\eqsecnum
\address{Ternopil State Technical University, Department of Physics,\\ 
56 Rus'ka Str., UA--46001, Ternopil, Ukraine; E-mail: vaha@tu.edu.te.ua}
\author{L.Didukh and V.Hankevych} 

\date{}
\title[Pressure-temperature phase diagram of generalized Hubbard model...]
{Pressure-temperature phase diagram of generalized Hubbard model with
correlated hopping at half-filling}

\begin{document}
\maketitle

\begin{abstract}
In the present paper pressure-temperature phase diagram of a 
generalized Hubbard model with correlated hopping in a paramagnetic state 
at half-filling is determined by means of the generalized mean-field 
approximation in the Green function technique. The constructed phase diagram 
describes metal-to-insulator transition with increasing temperature, and 
insulator-to-metal transition under the action of external pressure.
The phase diagram can explain paramagnetic region of the phase diagrams of 
some transition metal compounds.

\keywords{phase diagram, metal-insulator transition, correlated hopping}
\pacs{71.10.Fd, 71.30.+h, 71.27.+a}
\end{abstract}

\section{Introduction}

In the recent few years a generalized Hubbard model with correlated hopping
has been used widely to describe strongly correlated electron systems
(see papers~\cite{aa,dh1} and references therein); the electron-hole asymmetry
is a property of such a generalized Hubbard model as a result of the 
dependence of hopping integral on occupation of the sites involved in the 
hopping process. Recently, this model has been extended to the case of a 
doubly orbitally degenerate band~\cite{dsdh}.

The generalized Hubbard model has much richer properties than the well-known 
Hubbard model~\cite{hub}, and an usage of the electron-hole asymmetry 
conception allows to interpret the peculiarities of physical properties of 
narrow-band materials which are not explained by the Hubbard model.
In particular, the experimentally observed electron-hole asymmetries of
metal oxides conductivity, of cohesive energy of transition $3d$-metals and
of superconducting properties of high-temperature superconductors have been
explained within the generalized Hubbard model with correlated hopping
in papers~\cite{1_6,1_26,dhs}, \cite{1_6}, \cite{hir1,hir2,hir3} 
respectively.

Despite the fact that phase diagram of generalized Hubbard model has been
studied in works~\cite{1_16,1_20,1_51,1_63,1_64,1_65,1_79}, researchers 
pay no attention to a determination of the model phase diagram in a 
paramagnetic state under the action of external influences, in particular
pressure-temperature phase diagram. This task is related directly to the
problem of metal-insulator transition description under the action of 
external pressure and temperature, namely the constructed 
pressure-temperature phase diagram of the model would allow to describe
the observed metal-insulator transitions in narrow-band materials with
changing of pressure and temperature. An interest to such transitions 
is caused by the theoretical point of view as well as the rich possibilities 
of its application (see, for example, monograph~\cite{1_69} 
and review~\cite{3_23b}). Consequently, the goal of the present paper, being
a continuation of previous work~\cite{3_26} where the temperature-induced
metal-insulator transition was studied, is to determine the 
pressure-temperature
phase diagram of generalized Hubbard model with correlated hopping in a
paramagnetic state at half-filling. On basis of this phase diagram we
describe metal-insulator transition under the action of external 
pressure and temperature.

\section{Pressure-temperature phase diagram of the model}

Taking into account an external hydrostatic pressure $p$ we write the model 
Hamiltonian in the following form~\cite{1_6} (in this connection see 
also Ref.~\cite{3_32}):
\begin{eqnarray} \label{H_u}
H&=&-\mu \sum_{i\sigma}a_{i\sigma}^{+}a_{i\sigma}+
(1+\alpha u)t{\sum \limits_{ij\sigma}}'a_{i\sigma}^{+}a_{j\sigma}+
T_2{\sum \limits_{ij\sigma}}' \left(a_{i\sigma}^{+}a_{j\sigma}n_{i\bar \sigma}+h.c.\right)
\nonumber\\
&&+U \sum_{i}n_{i\uparrow}n_{i\downarrow} +{1\over 2}NV_0\kappa u^2,
\end{eqnarray}
where $i,\ j$ are the nearest-neighbours sites, $\mu$ is the chemical potential, 
$a_{i\sigma}^{+}, 
(a_{i\sigma})$ is the creation (destruction) operator of an electron of spin 
$\sigma$ ($\sigma=\uparrow, \downarrow$) on $i$-site  
(${\bar \sigma}$ denotes spin projection which is opposite to $\sigma$),
$n_{i\sigma}=a_{i\sigma}^{+}a_{i\sigma}$ is the number operator of electrons
of spin $\sigma$ on $i$-site,  $U$ is the intra-atomic Coulomb repulsion,
$t=t_0+T_1$, with $t_0$ being the matrix element of electron-ion interaction,
$T_1,\ T_2$ are the correlated hopping integrals (matrix elements of 
electron-electron interaction),
the primes at the sums in Hamiltonian~(2.1) signify that $i\neq j$.

The latest term of the Hamiltonian has meaning of the elastic energy of a 
uniformly deformed crystal, where $\kappa$ is the ``initial'' (purely lattice)
bulk elasticity, $N$ is the number of lattice sites, $u=\Delta V_0/V_0$ is
the relative change of the volume in unifrom strain ($V_0$ is the initial 
unit-cell volume). Formulating the Hamiltonian we have used the result of 
paper~\cite{3_32}: the dependence of a bandwidth $W$ on relative change of 
the volume $u$ in unifrom strain can be written in the form $W=2w(1+\alpha u)$, 
where $w=z|t|$ ($z$ is the number of nearest neighbours to a site), 
$\alpha = {V_0\over 2w}{\partial W\over \partial V}<0$. We assume
also that under the action of external pressure bandwidth changes only,
and the matrix elements of electron-electron interaction (the correlated 
hopping integrals and intra-atomic Coulomb repulsion) do not depend on
relative change of the volume.

As in papers~\cite{dh1,3_26}, using generalized mean-field 
approximation~\cite{1_6,did} (an analog of the projection operation) in the 
Green function method we obtain for a paramagnetic state at half-filling 
the energy gap width as
\begin{eqnarray} \label{en_gap_1}
&&\Delta E=-(1-2d)(w+\tilde{w})[1+\alpha u]+{1\over 2}(Q_1+Q_2),
\\
&&Q_1=\sqrt{\left[ B(w-\tilde{w})(1+\alpha u)-U\right]^2+[4dzt'(1+\alpha u)]^2},\\ 
&&Q_2=\sqrt{\left[ B(w-\tilde{w})(1+\alpha u)+U\right]^2+[4dzt'(1+\alpha u)]^2}, 
\end{eqnarray}
where $B=1-2d+4d^2,\ d$ is the concentration of polar states 
(holes or doublons) which has been calculated in Refs.~\cite{dh1,3_26}, 
$\tilde{w}=z|\tilde{t}|$, $\tilde{t}=t+2T_2,\ t'=t+T_2$; $t$ and $\tilde{t}$ 
are the terms describing hopping of quasiparticles within the lower 
and upper Hubbard bands (hopping of holes and doublons) respectively, $t'$
describes a quasiparticle hopping between hole and doublon bands (the processes 
of paired creation and destruction of holes and doublons).

According to the method proposed for the s(d)-f model in paper~\cite{3_35},
the equilibrium value of relative change of the volume $u$ is determined 
from the condition of minimum of the thermodynamic Gibbs' potential
\begin{eqnarray} \label{u1}
G=F+NpV_0(1+u),
\end{eqnarray}
where $F$ is the free energy. Using the known identity
$\partial F/\partial u=\langle\partial H/\partial u\rangle$, 
Eq.~(\ref{u1}) for the parameter $u$ can be represented as
\begin{eqnarray} 
\left\langle{\partial H\over\partial u}\right\rangle +NpV_0=0,
\end{eqnarray}
with $H$ being Hamiltonian~(\ref{H_u}). In the mean-field approximation 
passing to the space of quasi-momenta we get the following equation for
relative change of the volume $u$:
\begin{eqnarray} \label{u2}
\alpha u={2\alpha_1 V_0\over WN}\sum_{{\bf k}\sigma}t_{\bf k}\langle
a^{+}_{{\bf k}\sigma}a_{{\bf k}\sigma}\rangle +\tau pV_0,
\end{eqnarray}
where $\alpha_1$ is the parameter which determines the quantity 
$\partial W/\partial V,\ 2\alpha_1V_0/W\approx 0.1,\ 
\tau\approx 0.05$~eV$^{-1}$~\cite{3_32}.

Taking into consideration the fact that within the generalized mean-field 
approximation the first term of right-hand side of Eq.~(\ref{u2}) is
equal to zero~\cite{3_36} at the point of metal-insulator transition, 
we obtain the relation between relative change of the volume $u$ and 
an external hydrostatic pressure $p$ as
\begin{eqnarray} \label{u_p}
\alpha u=\tau pV_0.
\end{eqnarray}
Note that within the generalized Hartree-Fock approximation this equation 
is valid at the point of metal-insulator transition as well as in an
insulating phase.
 
To determine pressure-temperature phase diagram of the model we use 
formula~(\ref{en_gap_1}) for the energy gap width and the expression for
concentration of polar states calculated in Ref~\cite{3_26}.
Let us consider, for instance, the Mott-Hubbard compound NiS$_2$. This has
two electrons half filling an $e_g$ band, the 
half-width of initial (uncorrelated) band of this crystal is 
$w_0=z|t_0|\approx 1.05~$eV~\cite{3_8,3_27}, and the initial unit-cell
volume is $V_0\approx 14.79\times 10^{-30}~$m$^{3}$~\cite{3_37}. 
It shows the transition from the state of a paramagnetic insulator to the
paramagnetic metal state at a hydrostatic external pressure of 46 kbar and 
room temperatures. Thus the transition occurs for a decrease in volume of 
about 0.4\% with no change in crystal structure~\cite{3_6,3_7}. It also
becomes metallic on alloying with Ni$_2$Se, and the behaviour of this system
is discussed later in this section.

\begin{wrapfigure}[25]{l}{84mm}
%\begin{figure}
\begin{minipage}[b]{78mm} 
%\unitlength 1mm
%\begin{picture}(75,65)
%\put(0,66){\special{em:graph p_d_r.pcx}} 
%\end{picture}
\epsfxsize=75mm
\epsfysize=65mm
\epsfclipon
\epsffile{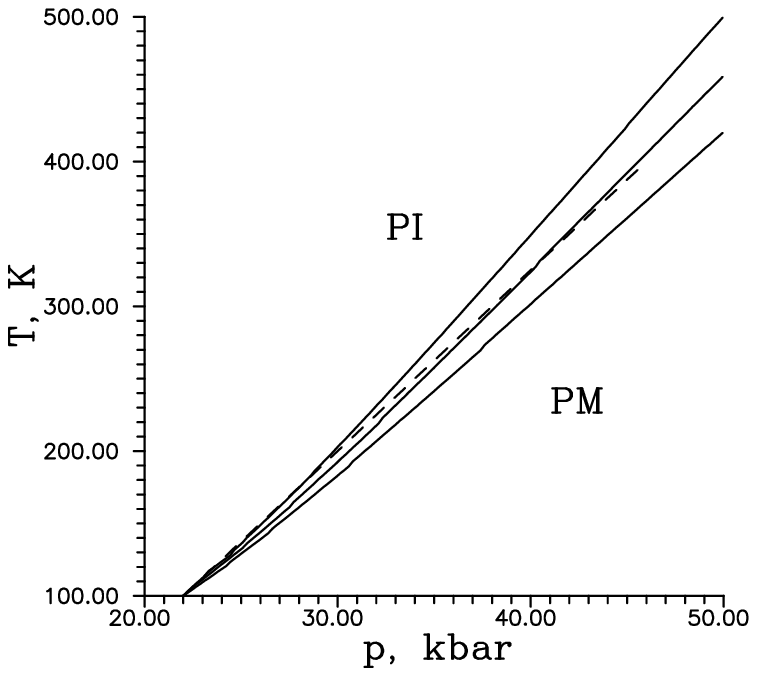}
\caption{}{The pressure-temperature phase diagram of metal-insulator transition
determined within generalized Hubbard model with correlated hopping
for NiS$_2$ in comparison with the experiment (the dashed curve):
$\tau_1=\tau_2=0$ (the upper curve); $\tau_1=0,\ \tau_2=0.1$ (the middle 
curve); $\tau_1=\tau_2=0.1$ (the lower curve). PI (PM) denotes paramagnetic
insulating (metallic) phase.} \label{p_d}
\end{minipage}
\end{wrapfigure}

To calculate the model parameter $U$ we fix one of the points 
($p=22~$kbar, $T=100$~K) of the experimental curve in the phase diagram (the 
dashed-line curve of figure~\ref{p_d}) and find the value of intra-atomic 
Coulomb repulsion $U$ at which the theoretical calculations within the
present model reproduce this point. Thus, we obtain:
${U\over w}\!=\!{U\over w_0}=2.0168$ for the correlated hopping parameters 
$\tau_1=T_1/|t_0|=0,\ \tau_2=T_2/|t_0|=0$ (these values of $\tau_1,\ \tau_2$ 
correspond to the Hubbard model), 
${U\over w}\!=\!{U\over w_0(1-\tau_1)}=1.79107$ for $\tau_1=\tau_2=0.1$, 
and ${U\over w}\!=\!{U\over w_0}=1.81437$ at $\tau_1=0,\ \tau_2=0.1$. 
Using these values of the model parameters we find the values of external
hydrostatic pressure and temperature at which energy gap width is equal to
zero (i.e. metal-insulator transition occurs).

The determined in this way pressure-tempe\-ra\-tu\-re phase diagram of the model 
(figure~\ref{p_d}) describes metal-insulator transitions in a paramagnetic 
state under the action of external pressure and temperature in NiS$_2$, 
namely the constructed phase diagram describes metal-to-insulator 
transition with increasing temperature, and insulator-to-metal transition 
under the action of external pressure. Comparison of this theoretically 
determined phase diagram with the phase diagram of the compound NiS$_2$ shows 
a good agreement between the theory and experiment. Besides, the theoretical 
calculations within the model reproduce the experimental data of 
paper~\cite{3_6} which point out a presence of the energy gap width $\Delta E>0$ 
in the ground state of NiS$_2$ and in absence of an external pressure.
The phase diagram shows that taking into account correlated hopping 
allows much better description of these experimental data than the Hubbard 
model; this testifies also much better physics of the present model and the 
important role of correlated hopping.

Analogous phase diagrams can be constructed for the other compounds:
(V$_{1-x}$Cr$_x$)$_2$O$_3$~\cite{1_69,3_5}, 
NiS$_{2-x}$Se$_x$~\cite{3_8,3_27} and Y$_{1-x}$Ca$_x$TiO$_3$~\cite{tok,3_29}
exhibiting such metal-insulator transitions. For example, the materials
(V$_{0.96}$Cr$_{0.04}$)$_2$O$_3$ shows a metal-insulator transition
at a hydrostatic external pressure of 13 kbar and room temperatures;
the transition occurs for a small decrease in volume of about 1\% with 
no change in crystal structure~\cite{3_5}. In (V$_{1-x}$M$_{x}$)$_2$O$_3$ 
(with $M=Cr,\ Ti$) the addition of Ti$^{3+}$ ions to V$_2$O$_3$ leads to
insulator-to-metal transition, whereas the addition of Cr$^{3+}$ ions results 
in metal-to-insulator transition. The simplest explanation is that the 
substitution of V$^{3+}$ ion for Cr$^{3+}$ ion leads to a band narrowing;
the Cr$^{3+}$ ion is a localized impurity and it deletes a state from the
$3d$-bands, deleting a state is equivalent to a band narrowing or an external
pressure decreasing~\cite{1_69} which drives the system towards the insulating
phase. Likewise the addition of Ti$^{3+}$ impurities is equivalent to an 
external pressure increasing. 

In the Mott-Hubbard compound NiS$_{2-x}$Se$_x$ 
electron hoppings between the sites of Ni occur by the chalcogenide sites
(this is caused by peculiarities of the pyrite crystal structure~\cite{kri}),
the substitution of S$^{2-}$ ion for Se$^{2-}$ ion in NiS$_2$ leads to 
an increase of wave functions overlapping, consequently the probability
of an electron hopping increases which is equivalent to a band broadening
or an external pressure increasing. Therefore, the pressure-temperature phase
diagram constructed for NiS$_2$ can describe the experimental
composition-temperature phase diagram~\cite{3_8,3_27} of the compound 
NiS$_{2-x}$Se$_x$.

Note also that to construct phase diagram of the system
Y$_{0.61}$Ca$_{0.39}$TiO$_3$ we have to generalize the previous results 
obtained at half-filling to the case of a non half-filled band because this
compound is characterized by such a band~\cite{3_29}.

In conclusion, in the present paper pressure-temperature phase diagram of the 
generalized Hubbard model with correlated hopping in a paramagnetic state 
at half-filling has been determined. The constructed phase diagram describes 
metal-to-insulator transition with increasing temperature, and 
insulator-to-metal transition under the action of external pressure. 
Comparison of this theoretically determined
phase diagram with experimental data, in particular with the phase diagram
of the compound NiS$_2$ shows a good agreement between the theory and 
experiment. 
We have found that taking into account correlated hopping allows much
better description of these experimental data than the Hubbard model; this
testifies also much better physics of the present model and the important
role of correlated hopping.

The determined pressure-temperarure phase diagram of the model can explain
paramagnetic region of the phase diagrams of the transition metal compounds: 
the systems NiS$_{2-x}$Se$_x$ and (V$_{1-x}$Cr$_x$)$_2$O$_3$, calcium doped 
YTiO$_3$.

\end{document}